\documentstyle[prl,amssymb,tighten,aps,multicol,epsfig]{revtex}

\begin{document}

\title{Improving Detectors Using Entangling Quantum Copiers}
\author{P. Deuar\cite{PDemail} and W. J. Munro\cite{WMemail}}
\address{Centre for Laser Science,Department of Physics, University of
Queensland, QLD 4072, Brisbane, Australia}
\date{\today}
\maketitle

\begin{abstract}
We present a detection scheme which using imperfect detectors, and
imperfect quantum copying machines (which entangle the copies), allows one to extract more
information from an incoming signal, than with the imperfect detectors
alone. 
\end{abstract}

\begin{multicols}{2}

   Copying machines in general use two approaches. One of the extreme cases
is a classical copying machine, where measurements (destructive or
non-destructive) are made on the original state, the results of which
are then fed as parameters into some state preparation scheme which
attempts to construct a copy of the original. This approach obviously
allows one to generate an arbitrary amount of copies, possibly all identical to
each other.  The opposite extreme is a fully quantum copying machine
which by some process that is unseen by external observers, creates a
fixed number of copies, usually destroying the original in the
process. Naturally in a realistic situation, noise will additionally
degrade the quality of the copies, and copiers which utilise both of the 
processes above are obviously also possible.

Ignoring for now the matter of the inevitable noise, the exact state
of the original can only be determined with certainty by some
measurement if all the possible states of the original  are mutually
orthogonal. In all other situations, any classical copying machine
must have a finite probability of producing imperfect copies. In fact,
by the well-known no-cloning theorem\cite{WZ:82,Barnumetal:96} 
the same can be said of quantum copying machines. If the possible
states of the original are not mutually orthogonal, there is no quantum
copier which will always make perfect copies. So one might ask what
good \emph{are} quantum copiers, then? 
Well, the obvious answer is that for the situation where the possible
originals are not orthogonal, often quantum copiers can  create better
copies than classical ones. Some examples are the UQCM for unknown
qubits\cite{BuzekH:96}, or other copiers for two non-orthogonal qubits\cite{Brussetal:98}.

While this promises the possibility of many applications of quantum
copying in the future, few specific examples of uses for a quantum
copier have been considered so far. When discussing practical
applications, quantum copiers have mainly been put forward as
something to be defended against by quantum cryptography
schemes. 
This article presents a analysis of a possible application
of quantum copiers: using them to improve detection efficiencies. 

\begin{figure}
\center{\epsfig{figure=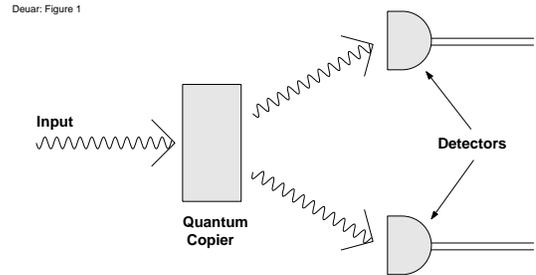,width=70mm}}
\caption{Basic Detection scheme using imperfect detectors, and a Quantum Copier}
\label{SCHEMEfig}
\end{figure}

We firstly note that in practice one always has restricted
detector resources. In particular, this article treats the situation
where the best available detectors have some efficiency less than
one. As an example system, consider the case where one of a set of possible input states are
to be distinguished by a measurement scheme, using (some number of
identical) imperfect detectors. One also has some (identical) 
quantum copiers which can act on the possible input states. 
At first, let us suppose that the possible input states are 
mutually orthogonal,
and that one has somehow acquired 
perfect quantum copiers for this set of states. Assume the copiers
destroy the original, and produce two copies for simplicity. Then, an obvious way
to take advantage of the copiers is to send the originals through a
quantum copier, before trying to detect both copies separately. 
(depicted in figure~\ref{SCHEMEfig}). This basically gives one a second chance to
distinguish the input state, if the detection at the first copy fails.

Consider a very simplified model of
photodetection using this measurement scheme. Suppose one has perfect
copiers, and noiseless photodetectors of efficiency $\eta$. That is, the probability
of a a count on the detector is $\eta$ if a photon is incident, and
$0$ otherwise.  With the copier set up as in figure~\ref{SCHEMEfig}, if any of the
detectors register a count, one can with certainty conclude that a
photon was incident. So, if a photon \emph{is} incident, the
probability of finding it is 
\begin{equation}
P^{(1)}_{\text{count}|\text{photon}} = 
\eta+(1-\eta)\eta
\end{equation}
as opposed to just $\eta$ with no copier,   because one gets a
``second chance'' at detection. On the other hand, if no count is 
registered, then the probability that no photon was incident is
\begin{equation}\label{Boo}
P^{(1)}_{\text{nophoton}|\text{nocount}} =
\frac{1-p}{1-\eta p(2-\eta)}
\end{equation}
where $p$ is the probability
that a photon is incident on average, irrespective of the measurement
result. The expression of equation\ (\ref{Boo}) is always greater than
$\frac{1-p}{1-\eta p}$, which is the probability if no copier is
used. This increase reflects the added confidence that comes from
both detectors failing to register the photon. 

We note that using quantum copiers, and not classical
ones is vital. A classical copier would have to rely on the same imperfect
photodetectors, and would actually \emph{reduce} the
detection efficiency, since to detect a photon at one of the two copy
detectors, one must have been first detected at the copier. This gives
$P^{(1)}_{\text{count}|\text{photon}} = \eta^2(2-\eta)$ which is
always less than or equal to $\eta$, a result achieved without
any copiers at all.

Detection with the help of perfect quantum copiers, as briefly
discussed above,  is all very well,
but what happens when the equipment used is noisy, and not 100\%
efficient?  Consider the following, more realistic, model of photodetection. 
The possible states that are to be distinguished are the vacuum
$\left| {0} \right\rangle$ and single photon $\left| 1 \right\rangle$ states. 
The \emph{a priori} probability that the input state is a photon is $p$. 
A generalised measurement on some state $\hat{\rho}$ can be modeled by a
positive operator-valued measure (POVM) $\{\hat{A}_i\}$ \cite{Kraus:83,CavesD:94}
described by a set of $n$ positive operators $\hat{A}_i$,
 such that $\sum_{i=1}^{n} \hat{A}_i = \hat{I}$, where $\hat{I}$ is the
identity matrix in the Hilbert space of $\hat{\rho}$ (and of the $\hat{A}_i$).  The probability of
obtaining the $i$th result, by measuring on a state $\hat{\rho}$ is then
\begin{equation}\label{Pi}
P_i =\mbox{Tr}\left[\hat{\rho}\hat{A}_i\right] 
\end{equation}

Now suppose the photodetectors at one's disposal are noisy and have quantum
efficiency $\eta$. The effect of these can be modeled by the POVM
\begin{mathletters}\label{DETpovm}
\begin{eqnarray}
  \hat{A}_+ =& \eta \left| 1 \right\rangle \left\langle 1 \right| + 
  \eta\xi\left| 0 \right\rangle\left\langle 0 \right|& \\
  \hat{A}_- =& (1-\eta)\left| 1 \right\rangle \left\langle 1 \right| +(1-\eta\xi)\left| 
  0 \right\rangle \left\langle 0 \right|&
\end{eqnarray}\end{mathletters}
where the operator $\hat{A}_+$ represents a count, and the operator
$\hat{A}_-$ the lack of one.
The parameter $\xi\in[0,1)$  controls the amount of noise. That is, $\xi\eta$ is the
probability that the photodetector registers a spurious (``dark'') count when no photon is
incident.

We will model the quantum copier as one which has a probability $\varepsilon$ of
working correctly and producing perfect copies. Otherwise, the parameter
$\mu\in[-1,1]$ determines (in a somewhat arbitrary way) what is produced. This
can be written
\begin{mathletters}\label{COPYtransf}\begin{eqnarray}
\hat{\rho}_1 = \left|1\right\rangle\left|d\right\rangle
\left\langle 1 \right|\left\langle d \right| &\to
\varepsilon\left|1\right\rangle\left|1\right\rangle\left\langle 1 
\right|\left\langle 1 \right| +(1-\varepsilon)\hat{\rho}_N = \hat{\rho}^1_1\\ 
\hat{\rho}_0 = \left|0\right\rangle\left|d\right\rangle \left\langle 0 \right|\left\langle d \right| &\to 
\varepsilon\left|0\right\rangle\left|0\right\rangle \left\langle 0 \right|\left\langle 0 \right| +
(1-\varepsilon)\hat{\rho}_N = \hat{\rho}^1_0
\end{eqnarray}\end{mathletters}
where $\left| d \right\rangle$ is a dummy state, which is fed into the copier, and becomes
the second copy. It is included here to preserve unitarity in the
perfect copying case $\varepsilon=1$. The state produced upon failure of the
copier, $\hat{\rho}_N$ is independent of the original, and is given
by
\begin{equation}\label{NOISEstate}
\hat{\rho}_N = (1-|\mu|)\frac{\hat{I}}{4} + \
\left\{ \begin{array}{cll}
\mu&\left|1\right\rangle\left|1\right\rangle\left\langle 1 
\right|\left\langle 1 \right| & \text{ if }\mu>0 \\
|\mu|&\left|0\right\rangle\left|0\right\rangle\left\langle 0 
\right|\left\langle 0 \right| & \text{ if }\mu\le0
\end{array}\right.
\end{equation}
Here, $\frac{1}{4}\hat{I}$ is the totally random mixed state.
 So, for $\mu=0$ a totally random noise
state is produced upon failure to copy, for $\mu=-1$ vacuum, for
$\mu=1$ photons in both copies, and for intermediate values of
$\mu$ a linear combination of the three cases mentioned.

This model (equation\ (\ref{COPYtransf})) of the copier is an extension (to allow for inefficiencies)
of the Wootters-Zurek copier, which has been extensively studied
\cite{WZ:82,BuzekH:96}. In the ideal case ($\varepsilon = 1$), with the
dummy input  state in the vacuum ($\left| d \right\rangle = \left| 0 \right\rangle$),  the
transformation is: 
\begin{equation}\label{cnott}
\left|0\right\rangle\left|0\right\rangle \to \left|0\right\rangle\left|0\right\rangle \qquad 
\left|1\right\rangle\left|0\right\rangle \to \left|1\right\rangle\left|1\right\rangle
\end{equation}
This transformation can be implemented by the simplest of all quantum logic circuits,
the single controlled-not gate. These have  recently begun to be  implemented
for some systems (although admittedly not for single-photon systems), and are the subject of intense ongoing research,
because of their application to quantum computing. This means that
similar schemes to the one considered here may become experimentally
realisable in the foreseeable future. We also point out that 
the transformation\ (\ref{cnott}) can be also considered an
``entangler'' rather than a copier. Consider its effect on the
photon-vacuum superposition state 
\begin{equation}
\frac{1}{\sqrt{2}}(\left| 0 \right\rangle +\left| 1 \right\rangle) \to 
\frac{1}{\sqrt{2}}(\left|0\right\rangle\left|0\right\rangle + \left|1\right\rangle\left|1\right\rangle) 
\end{equation}
This correlation between the copies is an essential property for the
detection scheme presented here to be useful --- otherwise one could
not combine the results of the different detector measurements to
better infer properties of the original. We will now examine how we 
determine whether the copying scheme we are proposing is more efficient.


Let us now consider the total amount of information about the input state that is contained in
the measurement results. This is the (Shannon) mutual information $I_m$ per input state
 between some observer $A$ who knows with certainty what the
 original states are (perhaps because they were prepared by that
observer), and another observer $B$ who has access to the measurement
results of the detection scheme.
This can be readily evaluated from the expression \cite{Shannon:48a,Shannon:48b,Hall:97}
\begin{equation}\label{Imexpr}
  I_m = \sum_{i,j} P_{j|i} P_i \log_2 \frac{P_{j|i}}{P_j}
\end{equation}
where $i$ ranges over the number of possible input states, and $j$
over the number of possible detection results.
$P_i$ are the \emph{a priori} probabilities that the $i$th input state
entered the detection scheme, $P_{j|i}$ is the probability that
the $j$th the detection result was obtained given that the $i$th state
was input, and $P_j$ is the marginal probability that the $j$th
detection result was obtained overall.

 This mutual information has very concrete meaning even though in general, $B$ can never be
actually certain what any particular input state was. It is known that
by using appropriate block-coding and  error-correction schemes, $A$ can transmit to $B$
an amount of \emph{certain} information that can come arbitrarily close to
the upper limit $I_m$ imposed by the detection probabilities. In other words,
$I_m$ is the maximum amount of
information that $A$ and $B$ can share using a given detection scheme,
if they are cunning enough. It follows then, that the detection scheme which gives a greater information
content about the initial state $I_m$, will be the potentially more useful one.
The authors have actually shown that
 the Wootters-Zurek copier is the optimal
quantum broadcaster of information when the information is decoded one-symbol at a
time \cite{Us}, and this will be discussed in a future paper.

From expression\ (\ref{Imexpr}) it can be seen that $I_m$ depends on the \emph{a priori} input
probabilities (the parameter $p$ in the cases considered here). This
leads one to surmise that (at least in general) various detection
schemes may do relatively better or worse depending on how frequently
the input is a photon. This is in fact found to be the case.
However, in what follows, we will concentrate mainly on the $p=\frac{1}{2}$ case of
equiprobable photons and vacuum, since
this is the situation which allows the maximum amount of information
to be encoded in the original message, and so is in some ways the most
basic case.

If the new detection scheme gives mutual information content
$I_m(\varepsilon,\eta,\mu,\xi,N,p)$ per input state, then  $\eta^e(I_m(\varepsilon,\eta,\mu,\xi,N,p))$ is defined
as the efficiency of a noiseless detector  that would give the same mutual
information content if it was used by itself in the basic scheme with no copiers. i.e.
\begin{equation}\label{nedef}
I_m(\cdot,\eta^e,\cdot,0,0,p) = I_m(\varepsilon,\eta,\mu,\xi,N,p)
\end{equation}
$\eta^e$ is a one-to-one, monotonically increasing function of
$I_m$, and so if (and only if) some detection scheme increases $\eta^e$, it also
increases the mutual information, thus $\eta^e$ and $I_m$ are
equivalent for ranking detection schemes in terms of effectiveness.
$\eta^e$ also has the advantage that for some cases of the new
copier-enhanced detection scheme it is
independent of the photon input probability $p$.

Now it is time to ask the question: For what parameter values does the
copier-enhanced detection scheme provide more information about the initial
states than using a single detector? 
Consider firstly the simplest case of interest, where there are no
spurious (dark) counts in the photodetectors ($\xi=0$), and one has a copier of efficiency
$\varepsilon$ which produces vacuum upon failure ($\mu = -1$). This will give
some idea about the relationship between the detector and copier
efficiencies required, leaving the effects of noise for later consideration.

As  mentioned previously, in
this situation the effective efficiency is independent of $p$, and
with one layer of copiers ($N=1$), it is found to be given by the simple
expression:
\begin{equation}\label{effeff}
\eta^e_{(1)} = \varepsilon\left[1-(1-\eta)^2\right]
\end{equation}
Since this is independent of $p$,  introducing a second lot of copiers,
is equivalent to  replacing $\eta$ in the above expression by
$\eta^e_{(1)}$ i.e. $\eta^e_{(n+1)} =
\varepsilon\left[1-(1-\eta^e_{(n)})^2\right]$. In fact, in the limit of
never-ending amounts of copiers, the effective efficiency approaches
\begin{equation}\label{ninf}
\lim_{N\to\infty} \eta^e  = 2 -\frac{1}{\varepsilon}
\end{equation}
One finds that effective efficiency
 is improved (over $\eta^e=\eta$) by the copier scheme whenever
\begin{equation}\label{simplecond}
\varepsilon > \frac{1}{2-\eta}
\end{equation}

Since no random noise is introduced by either copier or detector,
improvement is achieved whenever more copiers are added, to arbitrary
order $N$.   A few things of interest to note
\begin{itemize}
\item The copier efficiency required is always above $\eta$ and above
$\frac{1}{2}$.
\item A gain in efficiency can be achieved even with quite poor copiers
--- for relatively small detector efficiencies $\eta$ (which occur for
photodetection in practice), the copier efficiency required is only
slightly above half!
\item For very good detectors, to get improvement, the copier efficiency $\varepsilon$ has to be
slightly larger than the detector efficiency $\eta$.
\item For low efficiencies, the relative gain in efficiency can be
very high, and can reach approximately $2^N$ for very poor detectors
and very good copiers.
\end{itemize}

\begin{figure}
\center{\epsfig{figure=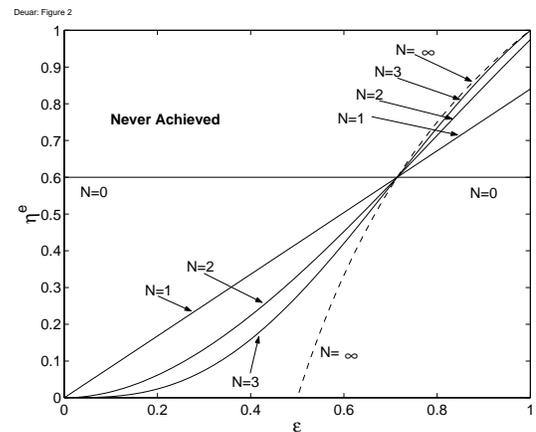,width=70mm}}
\caption{ Equivalent efficiency $\eta^e$ as a
function of copier efficiency $\varepsilon$ and number of levels of copiers
$N$ when detector efficiency is $\eta=0.6$, and both detectors and
copiers are noiseless ($\xi=0$, $\mu=-1$).
  Results for $N=0$ to $N=3$ are shown as solid lines, and the limit
of what can be achieved is shown as a dashed line. Regions beyond the
$N=0$ and $N\to\infty$ cases are not achievable with noiseless copiers.
The results are independent of the photon input  probability $p$, in this case.}
\label{eta0.3FIG}
\end{figure}

To examine how much improvement can be achieved in more detail,
consider when the efficiency of the detectors is $\eta=0.6$. This is a
typical efficiency for a pretty good single-photon detector at present. This
is shown by the solid lines in figure~\ref{eta0.3FIG}. Note how quite
large efficiency gains are achievable even when the copier efficiency
is slightly over the threshold useful value of $\epsilon=0.714$ (from
equation\ (\ref{simplecond})), and
how adding more copiers easily introduces more gains at first, but
after three levels of copiers, adding more becomes a lot of effort for not
much gain.

To conclude it can be seen that when one is restricted to
using imperfect detectors (as is always the case),
more efficiency of detection can be gained by employing entangling quantum
copiers such as a controlled-not gate. In fact if the efficiency of the detectors
is far from 100\% (such as in single-photon detection) the copier does not
have to be very efficient itself, and significant gains in detection
can still be made. We note that although a
detailed analysis was carried out for the case of single-photon detection,
the basic scheme can be readily generalised to other types of
detectors.

From\ (\ref{simplecond}), it can be
seen that to be useful, the quantum copiers must be successful with an
efficiency $\varepsilon$ over 50\% and somewhat greater than the detector
efficiency $\eta$. It is not generally clear how feasible this is for
 various physical systems, or measurement schemes that one might
wish to employ. With current technology it is often still
easier to make measurements on a system, rather than entangling it
with other known systems, however this varies from measurement to measurement
and from system to system. The physical processes involved in
measurement and quantum copying are often quite different: the former
requires creating a correlation between a quantum system and a
macroscopic pointer, whereas the latter involves creating quantum
entanglement between two similar microscopic states. Efficient
detection depends on correlating the system with its environment in a
strong, yet controlled way, whereas quantum copying depends on
isolating the system from its environment. One thus supposes that the
usefulness of a scheme such as the one outlined here will depend on
the system and measurements in question, due to the relative ease of
implementing detection and controlled quantum evolution in those systems.

WJM would like to acknowledge the support of the Australian Research Council.

\end{multicols}


\begin{references}
\bibitem{phovac}
For high detector efficiencies $\eta$, when dark counts
occur, photon default output upon failure to copy ($\mu=1$) may be
more desirable than vacuum.
\end{references}
\end{document}